\documentclass{iaus}
\usepackage{graphicx}

\title[HD~189733 transits \& eclipses] 
{Transits and secondary eclipses of HD~189733 with Spitzer}

\author[E.\ Agol et al.]   
{Eric Agol$^1$
Nicolas B.\ Cowan$^1$, James Bushong$^1$, Heather Knutson$^2$, David
Charbonneau$^2$, Drake Deming$^3$ \& Jason H.\ Steffen$^4$}

\affiliation{$^1$Dept. of Astronomy, Box 351580, University of Washington, Seattle, WA 98195, 
\\[\affilskip]
$^2$Harvard-Smithsonian Center for Astrophysics, 60 Garden St., Cambridge, MA 02138
\\[\affilskip]
$^3$NASA's Goddard Space Flight Center, Planetary Systems Laboratory,
Code 693, Greenbelt, MD 20771, USA
\\[\affilskip]
$^4$Fermilab Center for Particle Astrophysics, P.O. Box 500 MS 127, Batavia, IL 60510 }

\pubyear{2008}
\volume{253}  
\pagerange{}
\jname{IAU Symposium 253 Transiting Planets}
\editors{Queloz D., Sasselov D., Torres M.\ \& Holman M., eds.}

\begin{document}

\maketitle

\begin{abstract}
We present limits on transit timing variations and secondary eclipse depth
variations at 8 microns with the Spitzer Space Telescope IRAC camera. Due to
the weak limb darkening in the infrared and uninterrupted observing,
Spitzer provides the highest accuracy transit times for this bright system,
in principle providing sensitivity to secondary planets of Mars mass in
resonant orbits.  Finally, the transit data provides tighter
constraints on the wavelength-dependent atmospheric absorption by the planet.
\end{abstract}

\firstsection 

\section{Introduction}

The extremely precise 33 hour phase function measurement of HD~189733b
\cite[Knutson et al.\ (2007)]{Knutson2007} observed with the 8 micron IRAC camera on
the Spitzer Space Telescope yielded the most precise times of transit
and secondary eclipse for any extrasolar planet, 6 and 24 seconds, respectively.  
This led us to propose to observe an additional six transits and eclipses of this system 
over time with the goal of measuring:

\begin{itemize}
\item precise transit-timing (\cite[Agol et al.\ 2005]{Agol2005}, \cite[Holman \& Murray 2005]{Holman2005}) to
search for the presence of resonant (or near-resonant)
terrestrial-mass planets captured by migration (\cite[e.g.\ Mandell, Raymond \& Sigurdsson 2007]{Mandell2007})
or on longer timescales precession of an eccentric orbit (\cite[Miralda-Escud{\'e} 2002]{MiraldaEscude2002}, \cite[Heyl \& Gladman 2007]{Heyl2007}; also Fabrycky and Wolfe, these proceedings);
\item variations in the depth of the secondary eclipses with time which might be caused by large-scale variable
atmospheric features (\cite[Rauscher et al. 2007]{Rauscher2007}; also Showmand and Dobbs-Dixon, these proceedings);
\item precise transit depth for comparison with atmospheric-absorption models to contrain the molecular 
composition (\cite[e.g. Tinetti et al. 2007]{Tinetti2007}; also Tinetti et al., Fortney et al., and Hubeny et al., these proceedings);
\item improved system parameters for better characterization of the planet, host-star, and orbit properties (\cite[Winn et al.\ (2007)]{Winn2007}; also Winn, these proceedings).
\end{itemize}

For librating planets in a low-order mean motion resonance, the times of transit vary
with an amplitude:
\begin{equation}
\delta t_{2:1} \approx P_{trans} \left({M_{pert} \over M_{trans}}\right)
\approx 3 {\rm \ min} \left({P_{trans} \over 3 {\rm \ day}}\right) \left({M_{pert} \over M_{\oplus}}\right) \left({M_{J} \over M_{trans}}\right),
\end{equation}
and libration period of
\begin{equation}
P_{2:1} \approx P_{trans}\left({M_* \over M_{trans}}\right)^{2/3}
\approx 150 {\rm \ days}\left({P_{trans} \over 3 {\rm \ day}}\right) 
\left({M_{*} \over M_{\odot}}\right)^{2/3} 
\left({M_{J} \over M_{trans}}\right)^{2/3},
\end{equation}
where $M_\oplus, M_J, M_*$ are masses of the Earth, Jupiter, and
host star, $P_{trans}$ is the period of the transiting planet, and
the numbers have been estimated for the libration amplitude for
planets starting on circular orbits with exact commensurability 
\cite[(Agol et al.\ 2005)]{Agol2005}; the actual value depends on
the libration amplitude.
This timescale requires observations separated by months with $\approx$ 
seconds precision in timing, and in principle could be sensitive to sensitive 
to Mars mass planets.

\section{Data reduction summary}

So far Spitzer has observed four transits and four secondary eclipses of HD~189733 for 
this program with 44,000
exposures of 0.4 second each over 5 hours each.  Additionally we re-analyzed 
the data from \cite[Knutson et al.\ (2007)]{Knutson2007}.  We utilized the IRAC 
8 microns as it has been demonstrated to be the most stable IRAC band.  Due to 
the brightness of the host star we made the observations in sub-array mode.  We carried out 
aperture photometry with a 3.5 pixel radius.  
Due to the small limb-darkening, 
stable instrument (thanks to the Earth-trailing orbit of Spitzer which leads to 
stable thermal properties and no occultation of targets by the Earth, as occurs 
with HST) we obtained 0.5\% precision per 0.4 second exposure.  In the shot-noise 
limit we expect a 3-second precision for transit times.

\begin{figure}[b]
\begin{center}
 \includegraphics[width=13cm]{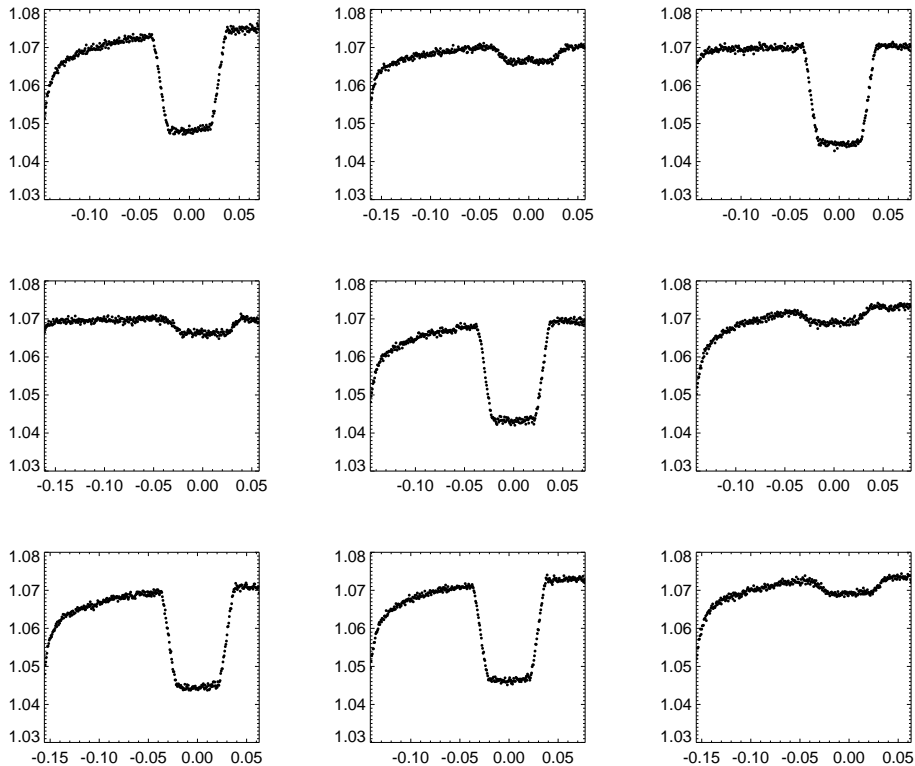} 
 \caption{Atlas of transits and secondary eclipses obtained at 8 microns
with Spitzer.  Horizontal axis is time in units of days; vertical axes
are photon counts in arbitrary units.}
   \label{fig01}
\end{center}
\end{figure}

Figure \ref{fig01} shows a gallery of the transits and secondary eclipses obtained.  
The strong ``ramp" which causes the flux to change by about 1\% is a well known 
feature of the IRAC camera when observing bright sources.  We fit the ramp with the 
function $a_0 + a_1 (t-t_s) + a_2 \ln{(t-t_s)}$, where $t_s$ is the start of the observation and 
$a_{0,1,2}$ are constants.  This appears to remove any trace of the ramp in the corrected 
data.  We assign each data point an error bar which equals the scatter in residuals 
of the data outside of transit/eclipse.

\section{Transit/eclipse depths and limb darkening}

\begin{figure}
\centering
\resizebox{6.5cm}{!}{\includegraphics{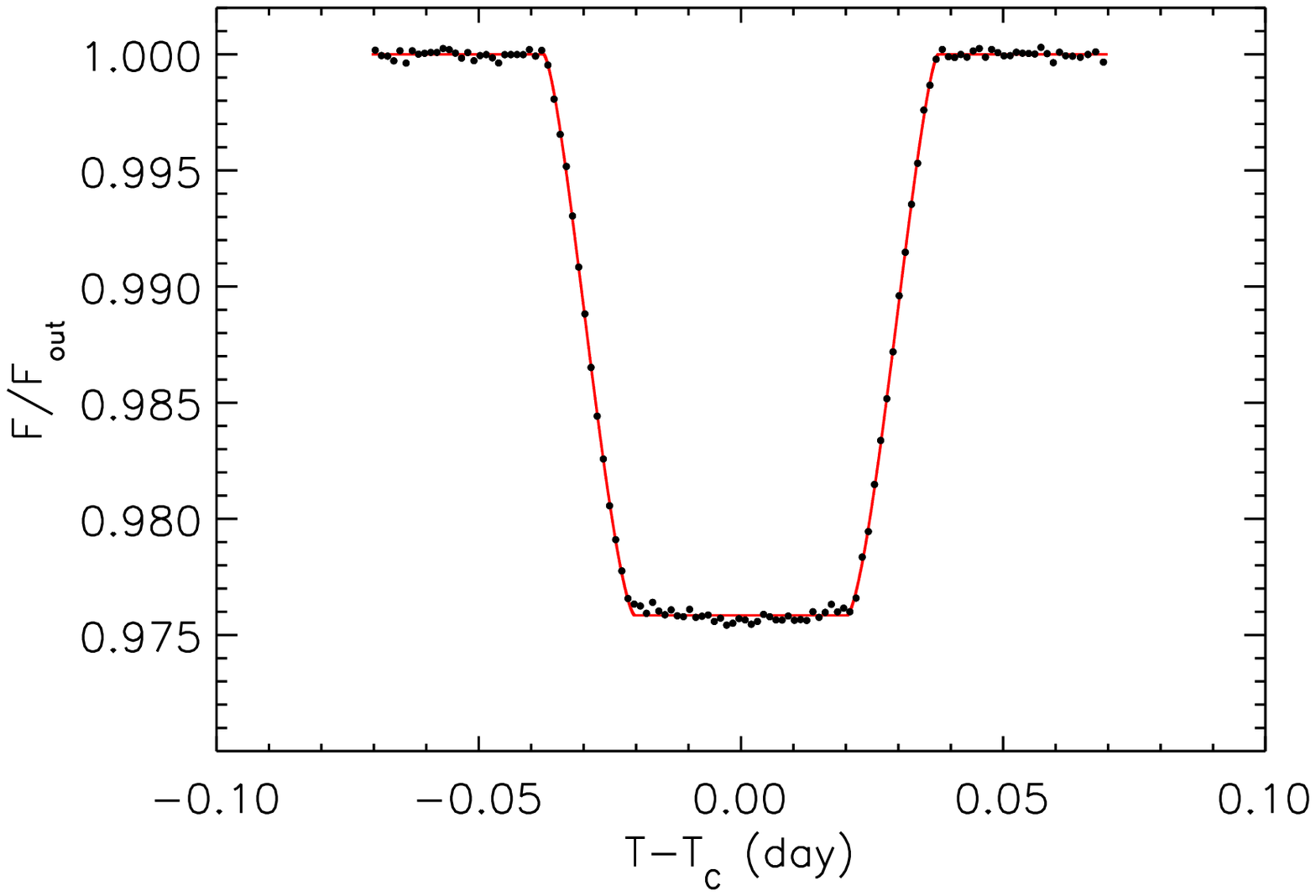} \label{fig02a}}
\resizebox{6.5cm}{!}{\includegraphics{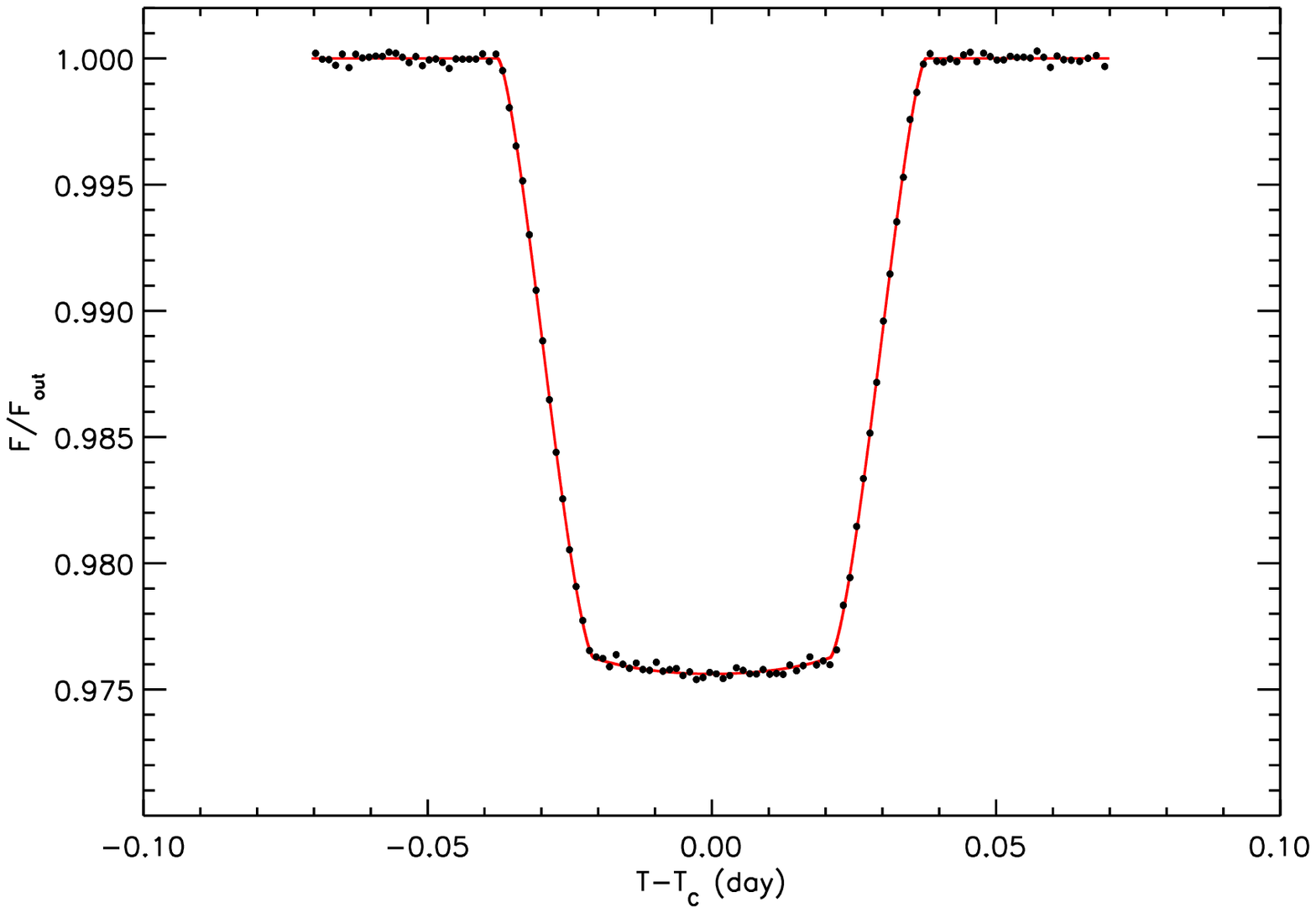} \label{fig02b}}
\caption[]{(a) Binned transits with best-fit model (solid red line) assuming no 
limb darkening for star.
Data include error bars. (b) Transits fit with best-fit model with limb darkening 
(red solid line).}
\end{figure}

\begin{figure}[b]
\begin{center}
 \includegraphics[width=13cm]{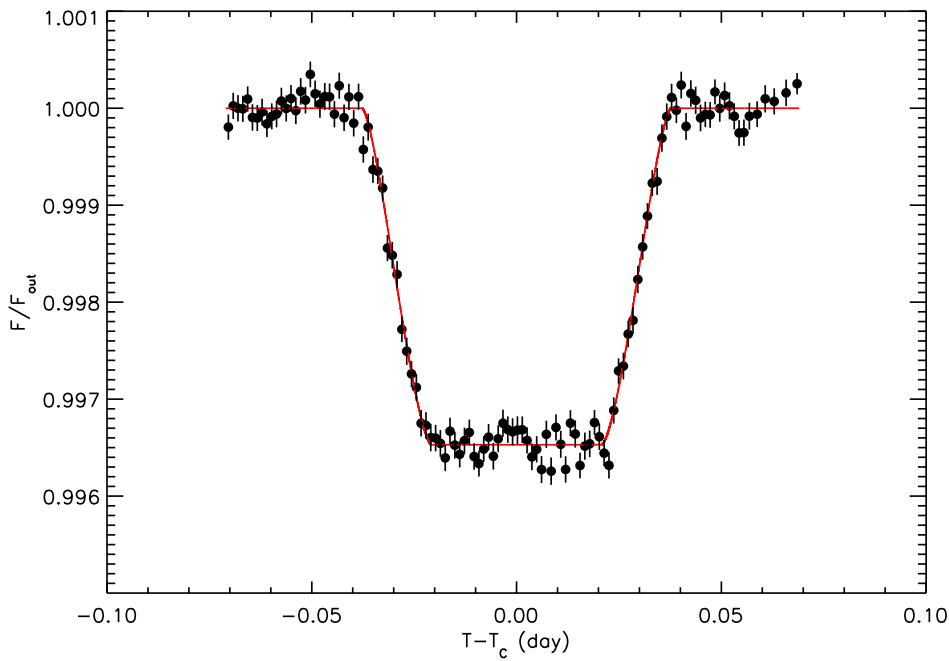}
 \caption{Binned data from 5 secondary eclipses with best-fit model (red solid line).}
   \label{fig02c}
\end{center}
\end{figure}

In two initial fits we include one ephemeris for the transits, one ephemeris for
the secondary eclipses, and required all other parameters describing the transits and 
eclipses to be the same for all 10 transits/eclipses to determine their
mean values.  We performed one of these fits with stellar limb-darkening set to zero, 
while for a second fit we allowed the linear coefficient of the stellar limb-darkening
to freely vary (in all cases we assume the planet to be uniform in surface brightness). 
We allowed the ramp parameters to vary independently for each transit/eclipse.
A sum of the five transits corrected for the ramp is shown in Figure \ref{fig02a}(a)
without limb darkening, and Figure \ref{fig02b}(b) with limb darkening.
The five secondary eclipses are shown in Figure \ref{fig02c}.

Without limb darkening we find a best-fit $\chi^2=232186.8$ for 231508
degrees of freedom (231516 data points with 8 model parameters).
With linear limb darkening the $\chi^2$ improves by $\Delta \chi^2= 140$
for 231507 degrees of freedom, with a best-fit limb darkening parameter
of $u_1 = 0.110 \pm 0.015$; thus limb-darkening is detected at 12$\sigma$ -
this can be seen by eye by comparing Figures \ref{fig02a}(a) and \ref{fig02b}(b).
This best-fit value for limb-darkening gives a limb-darkening 
profile which is very close to that predicted by a Kurucz model with parameters 
appropriate for HD~189733a (effective temperature $T_{eff} = 5000$ K and surface
gravity $log[g ({\rm cm/s^2})] = 4.5$).
We find a best-fit planet-star radius ratio of $R_p/R_* = 0.1558 \pm 0.0002$, which translates 
into a best-fit area ratio of $2.427\pm0.005$\%, while the best-fit eclipse depth is $0.347\pm 0.005$\% - note
that this is a 72$\sigma$ detection of a secondary eclipse!
The errors on these parameters (and throughout the rest of the paper) are computed 
by generating one hundred synthetic data sets from the best fit light curve (for each model) added to the residuals 
shifted by a random number of data points, re-fitting the model to each synthetic data set, and
then taking the standard deviation of the 100 synthetic parameter sets to determine
each parameter's error.

%

\section{Ephemeris and Transit Timing Variations}

We have measured the best-fit ephemeris to our data, separately for the transits and eclipses.  
We find a transit ephemeris of $T_0=  2454279.436741 \pm 0.000023$ HJD and $P=2.21857503 \pm 0.00000037$ days, while for the eclipses we find
$T_0=2454279.437407 \pm    0.000130$ HJD and $P=   2.21857306 \pm  0.00000209$ days, assuming that they occur exactly one half period after primary transit.
The periods are consistent within $1-\sigma$, while the central eclipse times differ by $57 \pm 11$ seconds, indicating that the secondary eclipses occur later.  About 31 seconds of this difference can be accounted for by the light travel time across the system, while the other 26 seconds is likely due to a slight orbital eccentricity.

\begin{figure}[b]
\begin{center}
 \includegraphics[width=13cm]{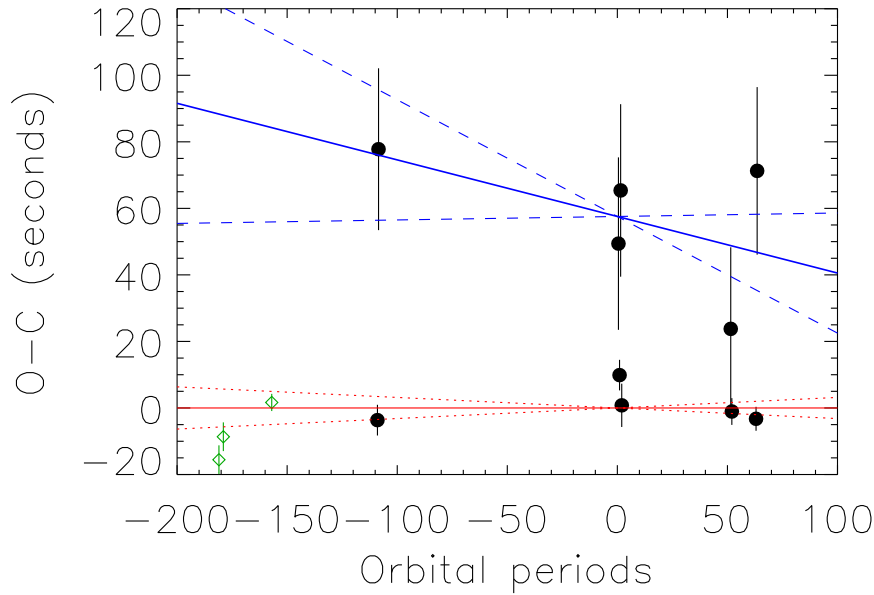} 
 \caption{Observed minus calculated (O-C) transit times for the five primary transits (black data points with small error bars) and the five secondary eclipses (black data points with larger error bars), compared to the ephemeris measured by fitting just the primary transits (red solid line) with uncertainties (red dotted lines).  The ephemeris from fitting the secondary eclipses is shown as a thick blue solid line, with dashed blue lines showing the uncertainty in slope.  Diamond green data points are transits from Pont et al.\ (2007) measured with Hubble Space Telescope with similarly sized error bars to our data.}
   \label{fig03}
\end{center}
\end{figure}

We have performed a second set of fits allowing the central times of transit/secondary eclipse to vary.  The transit timing variations are plotted in Figure \ref{fig03} for the transits and secondary eclipses, compared to the best-fit transit ephemeris (and assuming the secondary eclipse is offset by 1/2 an orbit).  We find no significant deviations from a uniform
period by more than 5 seconds for either the transits or secondary eclipses; once the final 
four data sets are obtained we will carry out a more detailed analysis.  Poster 79 by 
Miller-Ricci and poster 87 by Neale Gibson at this IAU symposium also present new 
transit-timing data for this system.

\section{Eclipse Depth Variation}

\begin{figure}[b]
\begin{center}
 \includegraphics[width=13cm]{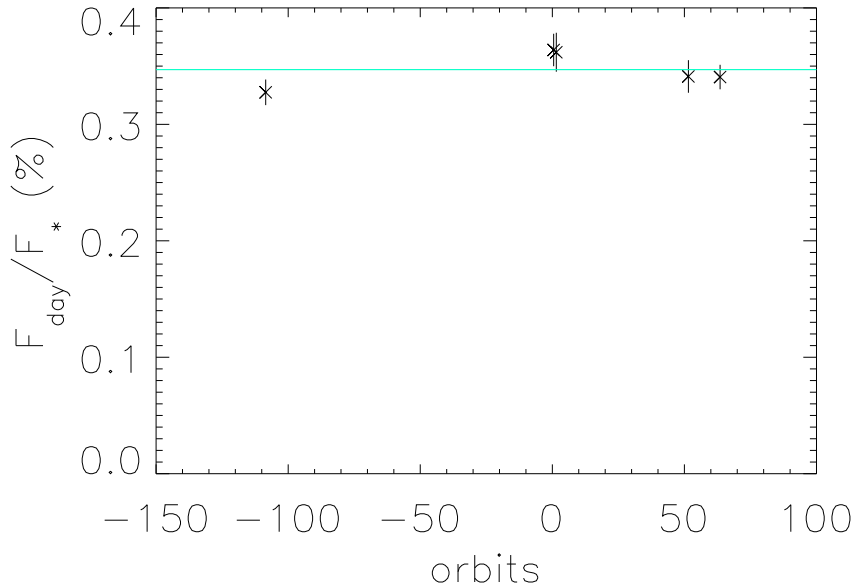}
 \caption{Changes in eclipse depths; cyan line is mean of 5 eclipses (0.347\%).}
   \label{fig04}
\end{center}
\end{figure}

In a third set of fits we held the ephemerides fixed, but allowed the depths of secondary eclipse to vary.  Figure \ref{fig04} shows the variations in the five observed eclipse depths in units of the flux of the star.  We find that the fractional variations in the eclipse depth are smaller than about 10\%; this is about the level of variation predicted by \cite[Rauscher et al.\ (2007)]{Rauscher2007}.  A fit to the eclipse depths assuming a constant depth gives a $\chi^2=6$ for 4 degrees of freedom (5 eclipses minus one free-parameter, the mean eclipse depth).  Once the final two secondary eclipse observations are obtained (in June and July 2008), a more comprehensive analysis will be carried out of the limits on secondary eclipse depth variation.

\section{Transit spectroscopy}

\begin{figure}[b] 
\begin{center}
 \includegraphics[width=13cm]{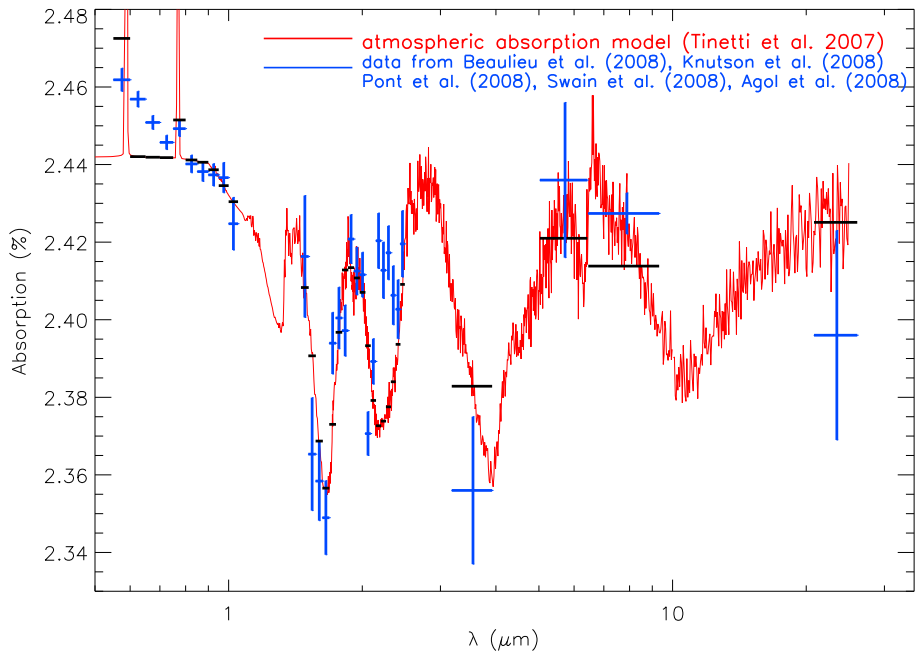}
 \caption{Compilation of data on transit depths of HD~189733.  Solid (red) curve is a
binned version of a model from Tinetti et al.\ (2007);  horizontal lines (black) are the mean
of this model over the bandwidths of each of the crosses (blue), which indicate 1-$\sigma$
error bars.}
   \label{fig05}
\end{center}
\end{figure}

The depth of primary transit we have obtained contains five times as much data as that in \cite[Knutson et al.\ (2007)]{Knutson2007}, providing more precise constraints on the spectral energy distribution in the mid-infrared.  We have combined this data point with transit depths measured by \cite[Pont et al.\ (2008)]{Pont2008} in the optical with HST, \cite[Swain et al.\ (2008)]{Swain2008} in the near-infrared with HST, \cite[Beaulieu et al.\ (2008)]{Beaulieu2008} at 3.6 and 5.8 microns with IRAC, and \cite[Knutson et al.\ (2008)]{Knutson2008} at 24 microns with MIPS (also Knutson et al., these proceedings).  The data are plotted in Figure \ref{fig05} along with the best-fitting model of \cite[Tinetti et al.\ (2007)]{Tinetti2007} which does not have enough methane to fit the near-infrared data.  We find that our 8 micron data point lies above the Tinetti model; as the 8 micron band covers a region in which methane absorption is quite strong ($\sim 10^{20}$ cm$^2$ per molecule), the model may be brought back into agreement with the data with a higher abundance of methane, as is already required by the near-infrared data (\cite[Swain et al.\ 2008]{Swain 2008}).  It also clear that the model is a poor fit to the optical data;  the fit may be improved by including Rayleigh scattering (Lecavelier des Etanges et al.\ 2008, and these proceedings).

\section{Conclusions}

The five transits we have observed are consistent with no transit timing variations
greater than 5 seconds.  The five secondary eclipses are consistent with no eclipse
depth variations at greater than the 10\% level.  From just an analysis of the
lightcurves we can place constraints on the eccentricity and longitude of periastron:
$\vert e \cos{\omega} \vert = 0.0002 \pm 0.0001$ and $\vert e \sin{\omega}\vert = 0.015
\pm 0.012$; similar to the results of \cite[Winn et al.\ (2007)]{Winn2007}.
We are collecting two more transits 
and two more secondary eclipses in Summer 2008; thus the results presented here are 
preliminary and require further more careful analysis.


\begin{thebibliography}{}

\bibitem[Agol et al.\ 2005]{Agol2005} {Agol, E., Steffen, J., Sari, R.\ \& Clarkson, W.} 2005,
 \textit{MNRAS} 359, 567
\bibitem[Beaulieu et al.\ (2007)]{Beaulieu2007} 
 {Beaulieu, J.~P., Carey, S., Ribas, I.\ \& Tinetti, G.}, 2008, \textit{ApJ}, 677, 1343
\bibitem[Heyl \& Gladman 2007]{Heyl2007} {{Heyl}, J.~S. and {Gladman}, B.~J.} 2007,
 \textit{MNRAS}, 377, 1511
\bibitem[Holman \& Murray 2005]{Holman2005}{Holman, M.~J.\ \& Murray, N.~W.} 2005,
 \textit{Science}, 307, 1288
\bibitem[Knutson et al.\ (2007)]{Knutson2007}
 {Knutson, H.S., Charbonneau, D., Allen, L.E., Fortney, J.J., Agol, E., Cowan, N.B.,
 Showman, A.P., Cooper, C.S.\ \& Megeath, S.T.} 2007, \textit{Nature}, 447, 183
\bibitem[Knutson et al.\ (2008)]{Knutson2008} {Knutson, H.~A., Charbonneau, D., Cowan, N.~B.,
	Agol, E., Showman, A.~P., Fortney, J.~J., Henry, G.~W.,
	Everett, M.~E.\ \& Allen, L.~E.} 2008, \textit{ApJ, submitted} 
\bibitem[Mandell, Raymond \& Sigurdsson 2007]{Mandell2007} {Mandell, A.~M., Raymond, S.~N.\ \& Sigurdsson, S.} 2007, \textit{ApJ}, 660, 823
\bibitem[Miralda-Escud{\'e} 2002]{MiraldaEscude2002}{Miralda-Escud{\'e}, J.},
 \textit{ApJ}, 2002, 564, 1019
\bibitem[Pont et al.\ (2007)]{Pont2007}{Pont, F., Gilliland, R.~L., Moutou, C., Charbonneau, D.,
	Bouchy, F., Brown, T.~M., Mayor, M., Queloz, D.,
	Santos, N.\ \& Udry, S.}, 2007, \textit{Astronomy \& Astrophysiscs}, 476, 1347
\bibitem[Pont et al.\ (2008)]{Pont2008}{Pont, F., Knutson, H., Gilliland, R.~L., Moutou, C.,
	Charbonneau, D.} 2008, \textit{MNRAS}, 385, 109
\bibitem[Rauscher et al. 2007]{Rauscher2007} {Rauscher, E., Menou, K., Cho, J.~Y.-K., Seager, S.\ \& 
 Hansen, B.~M.~S.}, 2007, \textit{ApJL}, 662, L115
\bibitem[Swain et al.\ 2008]{Swain2008}{Swain, M.~R., Vasisht, G.\ \& Tinetti, G.}, 2008,
 \textit{Nature}, 452, 329
\bibitem[Tinetti et al. 2007]{Tinetti2007} {Tinetti, G., Vidal-Madjar, A., Liang, M.-C., 
	Beaulieu, J.-P., Yung, Y., Carey, S., Barber, R.~J.,
	Tennyson, J., Ribas, I., Allard, N., Ballester, G.~E.,
	Sing, D.~K.\ \& Selsis, F.} 2007, \textit{Nature}, 448, 169
\bibitem[Winn et al. (2007)]{Winn2007}{Winn, J.~N., Holman, M.~J., Henry, G.~W., 
 Roussanova, A., Enya, K., Yoshii, Y., Shporer, A., Mazeh, T., Johnson, J.~A., 
 Narita, N.\ \& Suto, Y.} 2007, \textit{AJ}, 133, 1828
\end{thebibliography}
\end{document}